\documentclass[letter]{aa} 

\usepackage{graphicx}
\usepackage{txfonts}

\usepackage{ulem}

\begin{document}

   \title{The efficient photodesorption of nitric oxide (NO) ices }
\subtitle{A laboratory astrophysics study}
\titlerunning{Efficient photodesorption of NO ices}

   \author{R. Dupuy\inst{1},
                G. F\'{e}raud\inst{1},
                M. Bertin\inst{1},
                X. Michaut\inst{1},
                T. Putaud\inst{1},
                P. Jeseck\inst{1},
                L. Philippe\inst{1},
                C. Romanzin\inst{2},
                V. Baglin\inst{3},
                R. Cimino\inst{4}               
          \and
          J.-H. Fillion\inst{1}
          }
\authorrunning{Dupuy et al.}
   \institute{Laboratoire d'Etude du Rayonnement et de la Matière en Astrophysique et Atmosphères (LERMA), Sorbonne Universités, UPMC Univ. Paris 06, Observatoire de Paris, PSL Research University, CNRS, F-75005, Paris, France\\
        \and
        Laboratoire de Chimie Physique (LCP), CNRS UMR 8000, Univ. Paris Sud, F-91400 Orsay, France\\
        \and    
           CERN, CH-1211 Geneva 23, Switzerland\\
            \and
            Laboratori Nazionali di Frascati (LNF)-INFN I-00044 Frascati}         

   \date{}


 \abstract{The study and quantification of UV photon-induced desorption of frozen molecules furthers our understanding of the chemical evolution of cold interstellar regions. Nitric oxide (NO) is an important intermediate species in both gas-phase and solid-phase chemical networks. In this work, we present quantitative measurements of the photodesorption of a pure NO ice. We used the tunable monochromatic synchrotron light of the DESIRS beamline of the SOLEIL facility near Paris to irradiate NO ices in the 6 - 13.6 eV range and measured desorption by quadrupole mass spectrometry. We find that NO photodesorption is very efficient, its yield being around $10^{-2}$ molecule per incident photon for UV fields relevant to the diffuse and dense interstellar medium. We discuss the extrapolation of our results to an astrophysical context and we compare photodesorption of NO to previously studied molecules.} 


\maketitle

\titlerunning{R. Dupuy et al.: Photodesorption of nitric oxide ices}

\section{Introduction}

Nitric oxide (NO) is an important molecule in the interstellar medium (ISM) and is a key intermediate species from atomic N to molecular N$_2$, at the basis of the synthesis of nitrogen hydrides \citep{herbst1973,akyilmaz2007,hily-blant2010,legal2014}. Gas-phase NO has been observed in extremely varied environments: in photon dominated regions \citep{jansen1995}, in several molecular clouds \citep{ziurys1991,gerin1992,akyilmaz2007}, in comet P/Halley \citep{wallis1987}, in a circumstellar envelope \citep{quintana-lacaci2013}, and even in an extragalactic source \citep{martin2003}.
  
Unfortunately, there is no observational evidence for the existence of NO in the icy mantles of dust grains; however,  it has been reported several times that NO is highly likely to be present there.   Indeed, observations of gas-phase nitrogen-containing molecules in several starless cores gave a much lower  atomic N abundance  than the cosmic N abundance \citep{hily-blant2010}, so ices should contain N-bearing molecules (10 to 34~\% of total nitrogen could be contained in ices around protostars, see \cite{oberg2011}). A critical review of the nitrogen budget can be found in \citet{boogert2015}. The abundance of NO  in the interstellar solid phase is estimated at $\sim$~0.5 to 5~\% relative to H$_2$O \citep{charnley2001,theule2013,ruaud2016} i.e. within the same order of magnitude as the observed NH$_3$ or CH$_4$ ice abundances (e.g. \cite{oberg2016}).
Spatially resolved observations of gas-phase NO in pre-protostellar cores have shown that NO is depleted in the core \citep{akyilmaz2007},  i.e. there is a decrease in gas-phase NO in the core with respect to the edges. This depletion  can be explained by condensation of gas-phase NO onto icy grains \citep{hily-blant2010}, therefore enriching the outer ice with solid NO.   
In addition to gas-phase condensation, NO could also be directly synthesized in the icy mantle; laboratory studies have shown that solid NO can be produced by electron-irradiation of H$_2$O:N$_2$ or CO:N$_2$ ice mixtures \citep{kim2011,zheng2011}. 

It is commonly assumed that interstellar ices in dense cores are composed of two phases: an inner H$_2$O-rich phase containing NH$_3$, CH$_4$, and CO$_2$ and an outer CO-rich phase resulting from the gas-phase CO freeze-out on grains and containing CH$_3$OH and CO$_2$ \citep{boogert2015}. 
In comparison to the major components of astrophysical ices, solid NO is rarely considered even though it could play an important role in molecular clouds. Only a few gas-grain models include NO ices in these environments (e.g. \cite{charnley2001,ruaud2016}), whereas gas-phase NO is included in many models (e.g. \cite{legal2014}). There are also a few laboratory studies on NO ice chemistry and photoprocessing. In addition to NO formation in solid state \citep{kim2011,zheng2011}, it has been shown that solid state reaction of NO leads to more complex molecules such as NH$_2$OH and NO$_2$ \citep{congiu2012,minissale2014}. Moreover, hydrogenation and UV processing of ices containing NO in CO-, H$_2$CO-, or CH$_3$OH-rich ices lead to the formation of N-C containing molecules 
 \citep{fedoseev2016}, strengthening the importance of NO in the  solid-phase chemistry.

In the colder regions of the ISM where grains temperatures are below the sublimation temperatures of condensed molecules, the molecular gas-to-ice abundance ratio is believed to depend on the competition between accretion and non-thermal desorption processes \citep{bergin2007}. Among them, desorption stimulated by UV photons, so called UV photodesorption, can be an important mechanism \citep{Dominik2005,Hogerheijde2011,Willacy2000}.  
This has motivated a number of laboratory studies in order to quantify the efficiency of this desorption process for a collection of small molecules in their solid phase (H$_2$O, CO, N$_2$, CO$_2$), which have shown that photodesorption is a rather complex process at the molecular scale, with associated yields depending on the molecule, but also on many parameters such as ice composition, thickness, temperature, etc. (e.g. \cite{oberg2009,munozcaro2016}). 
A recent approach using tunable monochromatic light has been successful in providing photodesorption absolute yields in the vacuum ultraviolet (VUV) region (7-14 eV), highlighting the underlying photodesorption mechanisms and providing fundamental data that are independent from the energy profile of the irradiation source \citep{fayolle2011,bertin2013,fillion2014,dupuy2017}.

Among existing laboratory studies on the photodesorption of NO ice (for a review see \cite{zimmermann1995}), none provide absolute photodesorption yields at astrophysically relevant wavelengths. We present \textit{absolute} photodesorption spectra of pure NO ice between 6 and 13.6~eV (206.6 and 91.2~nm). This letter focuses on pure NO ice, as a first step, in comparison with another pure weakly bound species (CO) for which photodesorption spectra have been already measured in the same conditions.

\section{Methods}

Experiments are performed in the  Surface Processes \& ICES 2 (SPICES 2) set-up, a recent upgrade of the SPICES set-up of LERMA - Univ. Pierre et Marie Curie (France), which is especially designed for the high sensitivity and quantitative measurement of the photodesorption of neutrals and ions from molecular ices. It consists of an ultra-high vacuum (UHV) chamber with a base pressure of typically $10^{-10}$ mbar, within which a polycrystalline gold surface, a polycrystalline oxygen-free high conductivity (OFHC) copper surface, and a highly oriented pyrolytic graphite (HOPG) surface are mounted on a rotatable cold head that can be cooled down to \textasciitilde 10~K using a closed-cycle helium cryostat. Ices of nitric oxide NO (Airliquide, >99.9\% purity) are grown {in situ} by exposing the HOPG cold surface (10~K) to a partial pressure of gas using a tube positioned a few  millimetres  from the surface, allowing rapid growth without increasing the chamber pressure to more than a few 10$^{-9}$ mbar. Ice thicknesses are controlled with a precision better than 1 monolayer (1~ML corresponding to a surface molecular density of $\sim1\times10^{15}$ cm$^{-2}$) {via} a calibration procedure using the temperature programmed desorption (TPD) technique, as detailed in \citet{doronin2015}. 

The chamber is coupled to the undulator-based DESIRS beamline \citep{nahon2012} at the SOLEIL synchrotron facility, which provides monochromatic, tunable VUV light for irradiation of our ice samples. The coupling is window-free to prevent  cut-off of the higher energy photons. To acquire photodesorption spectra, the synchrotron light from a normal-incidence grating monochromator (narrow bandwidth around 25~meV or 0.3~nm) is scanned between 6~eV (206.6~nm) and 13.6~eV (91.2~nm). Higher harmonics of the undulator are suppressed using argon and xenon gas filters. Photon fluxes are measured upstream of the SPICES 2 set-up with a calibrated AXUV photodiode. The flux depends on the photon energy, and varies between 1.1 $\times$ 10$^{12}$ photons.cm$^{-2}$.s$^{-1}$ at 6~eV, 8.9 $\times$ 10$^{12}$ photons.cm$^{-2}$.s$^{-1}$ at 8~eV, and 4.6 $\times$ 10$^{12}$ photons.cm$^{-2}$.s$^{-1}$ at 13.6~eV.

During the photon energy scan, the relative amount of photodesorbed NO molecules is recorded by a quadrupole mass spectrometer (QMS) from the Pfeiffer company, with a mass resolution of $\sim$ 0.3 amu, as taken from the full width at half maximum of the mass peaks. A typical scan lasts around 3 minutes, which corresponds to a total photon fluence of at most  $\sim$ 10$^{15}$ photons.cm$^{-2}$.  We took care to limit the total fluence in each scan in order to get reproducible spectra, which are independent of the photon fluence. Under these conditions,  N or O desorption from pure NO ice could not be detected, either because their photodesorption is ineffective, or because the related photodesorption signals fall below our sensitivity limit, mainly given by our signal-to-noise ratio. 

Once corrected for the background contribution and divided by the photon flux, the QMS signal corresponds to the relative efficiency of photodesorption as a function of photon energy. The absolute photodesorption yield $y(\lambda)$ (in molecule per incident photon) is then deduced from that of CO \citep{fayolle2011} following a calibration procedure that takes into account NO electron impact ionization cross section in the ionization chamber of the QMS together with the apparatus function of the mass filter (see e.g. \cite{dupuy2017}). 
While the uncertainties on the relative photodesorption spectra are only due to experimental noise, background substraction, and photon flux uncertainties, this calibration method introduces much larger uncertainties on the absolute photodesorption yields. We estimate a systematic 40~\% uncertainty on the absolute photodesorption yields. The main source of error is the original calibration procedure of CO desorption yields in \citet{fayolle2011}; the other sources of error (flux, apparatus function, electron-impact ionization cross sections) are negligible in comparison.

 \section{Results and discussion}


\begin{figure}
    \resizebox{\hsize}{!}{\includegraphics{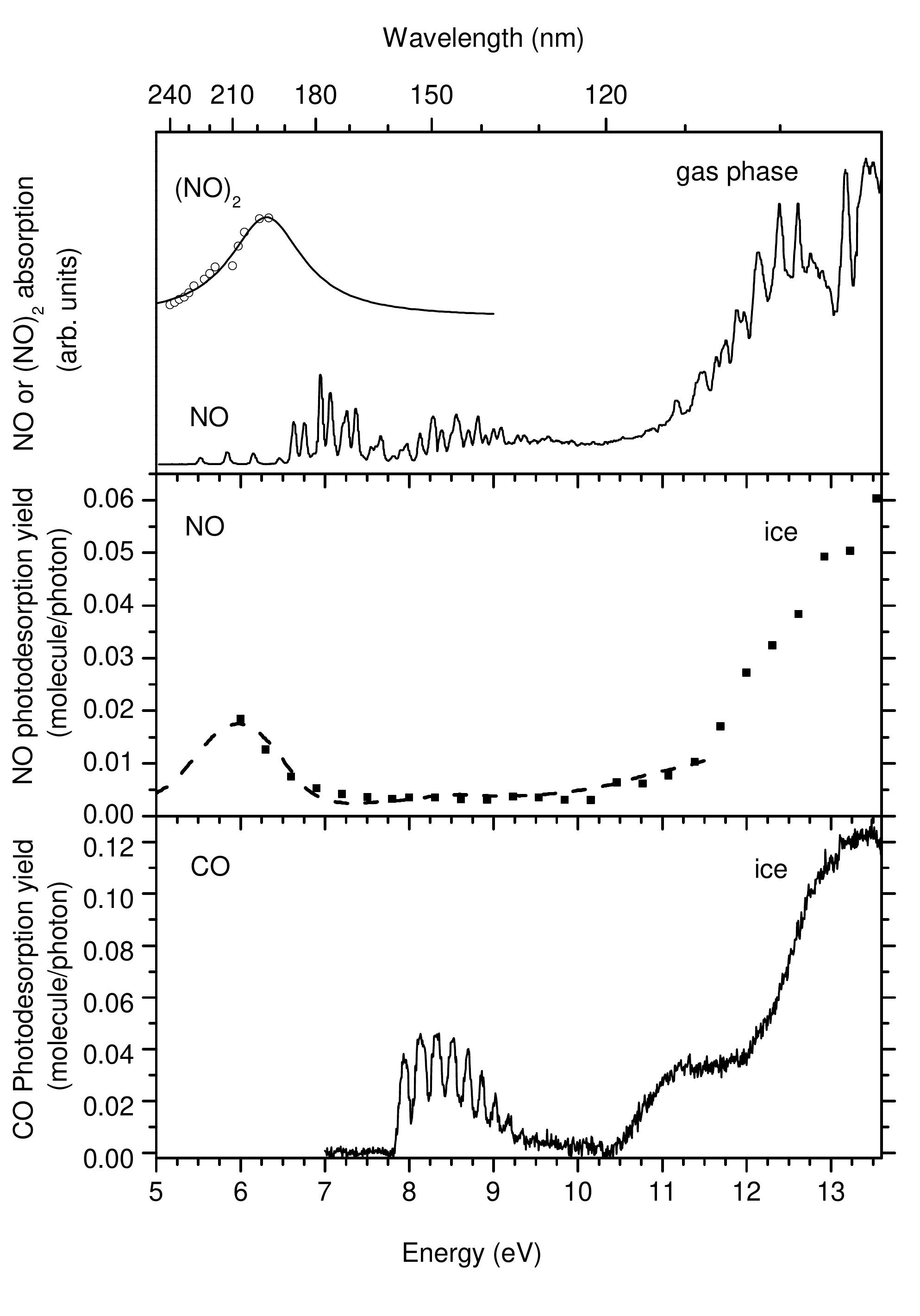}}
    \caption{\textit{Upper panel:} Electronic spectrum of gas-phase NO from \citet{chang1993} and electronic spectrum of gas-phase (NO)$_2$  from \citet{forte1978} (open dots) and its Lorentzian fit (solid line).
    \textit{Middle panel:} NO photodesorption spectrum of  10 ML thick NO ice on HOPG at 10~K between 6 and 13.6~eV (filled squares). Large energy steps of 0.3~eV were used to avoid ice ageing. Absorption spectrum of pure solid NO from \citet{lu2008} between 5 and 11.7~eV (dashed line). The absorption spectrum is scaled to the photodesorption spectrum.
     \textit{Lower panel:} CO photodesorption spectrum of a 20 ML thick CO ice on gold at 10~K between 7 and 13.6~eV \citep{dupuy2017}}
    \label{fig1}
\end{figure}

The NO photodesorption spectrum of 10 ML thick NO ice deposited on HOPG at 10~K is shown in Figure~\ref{fig1}. The photodesorption yield is absolute, representing the number of NO molecules desorbed per incident photon on the ice. Photodesorption of NO is found  over the whole 6 - 13.6~eV energy range. The photodesorption yield changes as a function of wavelength: there is a maximum around 6~eV, then it is very weak from 7~eV to the Lyman $\alpha$ (121.6 nm, 10.2 eV). It reaches a threshold around 10.5~eV (118~nm) and increases strongly for higher energies until it attains 0.06 molecule/photon at 13.6~eV.
The desorption spectrum looks like the absorption spectrum of solid NO from \cite{lu2008} recorded between 5 and 11.5~eV. This striking similitude between the absorption spectrum of the condensed molecule and the photodesorption spectrum has been already observed in a number of systems, and is the signature that the electronic excitation responsible for the desorption takes place in the condensed molecular film, and not in the supporting substrate.  

The assignment of the involved excited electronic states is ambiguous, especially as NO molecules are known to organize into dimer structures in the solid phase \citep{dulmage1953}, but the comparison between solid-phase and gas-phase NO and (NO)$_2$ spectra from the literature gives useful insights (Figure~\ref{fig1}). Unfortunately, we could not find the gas-phase absorption spectrum of (NO)$_2$ at energies higher than 6~eV. 
Gas-phase electronic spectra of NO and of its dimer are totally different in the 6~eV region; the absorption is very weak and well structured for NO, whereas a very broad band with a maximum at 6~eV is observed for (NO)$_2$ \citep{forte1978} that has been assigned to a charge transfer state \citep{Levchenko2006}. The state with a maximum at 6~eV  observed in absorption or in photodesorption spectra from solid NO can thus be attributed to the electronic excitation of (NO)$_2$. The binding energy of NO dimers is not very strong (0.086~eV or 1.99 kcal.mol$^{-1}$ from \citet{wade2002}); therefore, the desorbing molecules are NO monomers and not dimers.    
For energies above 10.5~eV, the strong increase in the desorption signal could be due to upper electronic states of NO or of (NO)$_2$, mainly assigned to Rydberg series that are clearly visible in the NO gas-phase spectrum of Figure~\ref{fig1} (\cite{chang1993} and references therein).
It could be argued  that photoelectrons produced by NO ionization are also at the origin of the measured desorption signal above 10.5~eV.  
However, the ionization energy of solid NO is measured at around 7.8~eV \citep{bertolo1990}, and electron-induced desorption experiments have shown that NO molecules are desorbed even by 0 eV electrons, i.e. there is no desorption threshold. In our experiment, desorption does not increase above 7.8~eV; therefore, photoelectrons produced in the ice should not play a major role in the desorption processes in this energy range.    

In the case of simple molecules such as CO and N$_2$, it has been shown that photodesorption is initiated by an electronic transition to a bound excited state, the relaxation of which subsequently leads to the desorption of a surface molecule. This process is referred to as an  indirect desorption induced by electronic transition (DIET) process because the energy can be transferred from the excited molecule to the surrounding species \citep{bertin2012, bertin2013}. For other molecules (CO$_2$, O$_2$, H$_2$O, CH$_3$OH), photodesorption can also be triggered by photodissociation of the molecule, followed by exothermic chemical recombination of the photofragments in the ice \citep{fillion2014, fayolle2013, bertin2016,martin-domenech2016,cruz-diaz2016} or momentum transfer between a photofragment and an intact molecule, also known as the kick-out mechanism \citep{Andersson2011}. In the case of NO, we have no evidence from our experiments that N or O atoms undergo photodesorption. The absence of efficient photodesorption of these light photofragments tends to rule out a potential kick-out mechanism. We cannot neglect photodesorption by photodissociation and exothermic chemical recombination of N and O atoms, i.e. reformation of NO molecules that can have enough energy to desorb. It is therefore difficult from our data set to conclude whether photodesorption is triggered by photochemistry, by an indirect DIET process, or both as  has been observed in the case of solid CO$_2$ \citep{fillion2014}. Finally, reactions between N or O fragments and intact NO are possible and could lead to the formation of bigger species, such as NO$_2$,
 during the irradiation.

It is interesting to compare NO photodesorption with that of the previously studied CO ice \citep{fayolle2011}. It is clear from Fig.~\ref{fig1} that photodesorption spectra of pure NO and  pure CO are very different in the 6-9.5~eV region. Photodesorption of NO  is a structureless broad band with a maximum at 6~eV, whereas that of CO shows vibronic structures spanning from 8 to 9.5~eV. Both NO and CO desorb very weakly at Lyman-$\alpha$, whereas above 10.5~eV the yields strongly increase. At high energy (> 10.5 eV), CO photodesorption yields are larger in \citet{dupuy2017} than in \citet{fayolle2011}, which we associate with an effect of the ice thickness, contrary to the lower energy part (7-10.5 eV) which has been found almost independent of the thickness for ices thicker than 3 ML \citep{bertin2012}. This is an interesting effect, that goes beyond the scope of this article, and we plan to study it more thoroughly in the future.

The average photodesorption yield Y$_{avg}$ for a given environment is derived using the  formula
$$ Y_{avg} = \frac{\int y(\lambda)\phi(\lambda)d\lambda}{\int\phi(\lambda)d\lambda}~, $$ 
where $y(\lambda)$ is the experimental photodesorption yield (Fig.~\ref{fig1}), and $\phi(\lambda)$ is the UV field of the environment.
NO photodesorption is $ Y_{avg} = 1.3 \pm 0.5 \times 10^{-2}$ molecule/photon for an inner core radiation field, and $Y_{avg} = 1.1 \pm 0.4 \times 10^{-2}$ molecule/photon for the interstellar radiation field (ISRF) (see  Sect. 2 for an estimation of the uncertainties). Two typical UV fields were used: the ISRF from \citet{mathis1983} relevant for the diffuse interstellar medium or at the edges of dense molecular clouds \citep{Heays2017}, and inner cloud fields generated by the interaction between cosmic rays and H$_2$ \citep{gredel1987}. Pure NO and pure CO photodesorption yields are of the same order of magnitude, both in dense core and in ISRF conditions.
The very high yields from pure NO ices appear to be typical of weakly interacting diatomic molecules.

It is clear from the absorption spectrum of solid NO from \citet{lu2008} that NO should still be desorbed by photon energies below 6 eV, which are missing in our experimental data due to technical limitations. However, if we extrapolate our photodesorption yields between 5 and 6 eV using the shape of the absorption spectrum, we find similar photodesorption yields within experimental errors for ISRF conditions. Indeed, half of the NO photodesorption yield comes from 12-13.6 eV photons.

Some insights on NO presence in different phases of ices can be deduced from laboratory experiments and observations. NO is a volatile species, an intermediary case between very volatile molecules such as O$_2$, CO, or N$_2$ and heavier molecules such as H$_2$O or CO$_2$.  
Spatially resolved observations of NO and N$_2$H$^+$ in dense cores have shown that NO and N$_2$ depletions occur at the dense core centre \citep{akyilmaz2007,pagani2012}, implying that NO and N$_2$ could coexist in the outer layer of the grain. Furthermore, experiments demonstrate that solid NO can be formed by energetic processing of H$_2$O:N$_2$ or  CO:N$_2$ ices \citep{kim2011,zheng2011}, which suggest that it could also be  present in the H$_2$O-rich phase. Thus, NO could be contained both in the outer layer with CO and within the H$_2$O-rich mantle, and this could be included in gas/grains models. The present pure phase investigation can be viewed as a preliminary study in the case of interstellar NO as photodesorption yields are certainly different for more complex systems such as NO:CO layers and mixtures, and NO:H$_2$O layers and mixtures in relevant abundances.
Indeed, NO dimerization could be hindered by other organizations in the case of CO:NO or NO:H$_2$O co-adsorption. Moreover, the effect of the molecular environment on desorption can drastically change the photodesorption yields. For instance, it has been shown that the photodesorption of some species co-adsorbed as mixes or layers in CO-rich ices is mainly due to the absorption of CO molecules via the indirect DIET mechanism. However, the efficiency of this process depends  on the nature of the species: it is very efficient for N$_2$ \citep{bertin2013}, becomes less effective for slightly more complex molecules such as CH$_4$ or CO$_2$ \citep{fillion2014, dupuy2017}, and finally is not observed in the case of CH$_3$OH \citep{bertin2016}. We therefore expect that the photodesorption yield of NO will be different in the case of realistic mixtures in CO- or H$_2$O-rich ices.
                                             
\section{Conclusions}  

We measured the NO photodesorption spectrum from pure NO ices and some of its features were assigned. Photodesorption of NO  from pure NO ices is very efficient, of the same order of magnitude as photodesorption yields of CO from a pure CO ice. The possible mechanisms at the origin of the photodesorption, photochemistry, and/or indirect DIET were also discussed. The high photodesorption efficiency observed for the pure NO ice system encourages a more detailed laboratory study on NO in representative molecular mixtures which include CO- and H$_2$O-rich ice phases, and we suggest that it could be included in gas-grain models. More laboratory studies, modelling, and perhaps observations of NO ices are needed to fill the NO-man's-land.

\begin{acknowledgements}
     
We thank L. Pagani for fruitful discussions concerning NO abundances in dense cores. This work was supported by the Programme National “Physique et
Chimie du Milieu Interstellaire” (PCMI) of CNRS/INSU with INC/INP co-funded by CEA and CNES, and by the European Organization for Nuclear Research (CERN) under the collaboration agreement KE3324/TE. We acknowledge SOLEIL for provision of synchrotron radiation facilities under the project 20150760 and we would like to thank Laurent Nahon and the DESIRS staff for assistance on the beamline. Financial support from the LabEx MiChem, part of the French state funds managed by the ANR within the investissements d'avenir program under reference ANR-11-10EX-0004-02 is acknowledged. Fundings by the Ile-de-France region DIM ACAV program is gratefully acknowledged.

\end{acknowledgements}

\bibliographystyle{aa} 
\bibliography{biblio_NO} 
%

\end{document}